# In-Ear Electrode EEG for Practical SSVEP BCI

Surej Mouli, Ramaswamy Palaniappan, Emmanuel Molefi and Ian McLoughlin

**Abstract:** Steady State Visual Evoked Potential (SSVEP) methods for brain–computer interfaces (BCI) are popular due to higher information transfer rate and easier setup with minimal training, compared to alternative methods. With precisely generated visual stimulus frequency, it is possible to translate brain signals into external actions or signals. Traditionally, SSVEP data is collected from the occipital region using electrodes with or without gel, normally mounted on a head cap. In this experimental study, we develop an in-ear electrode to collect SSVEP data for four different flicker frequencies and compare against occipital scalp electrode data. Data from five participants demonstrates the feasibility of in-ear electrode based SSVEP, significantly enhancing the practicability of wearable BCI applications.

## 1. Introduction

With the advent of the Internet of Things (IoT), wearable technology is rapidly evolving for uses such as physiological and biomechanical monitoring, combining data from multiple real-time sensors. Enhancing such technology with brain–computer interface (BCI) concepts has further potential for healthcare applications ranging from monitoring emotions, stress or other visuomotor tracking in real-time [1–3]. BCI applications already provide innovative solutions for various types of patients to overcome difficulties [4–7]. There are several non-invasive BCI modalities like P300, motor imagery (MI) and steady state visual evoked potential (SSVEP). Higher information transfer rate (ITR), minimal training requirement and better accuracy have led to SSVEP becoming the most popular BCI approach [8–10]. SSVEP is a natural response to a visual stimulus flickering at a constant rate, generating brain signals that oscillate approximately at the same frequency as that of the stimulus [11,12]. Current SSVEP based BCI applications collect data from the visual cortex using either dry or gel-based electrodes fitted on an electroencephalogram (EEG) cap. For a wearable BCI, usability is a prime concern.

Portable EEG devices like EMOTIV or StarStim which uses wet/dry electrodes need to be setup each time (10–15 min) when signal acquisition is required. Electrode contact quality also deteriorates with time and needs to be constantly monitored for signal quality. For practical SSVEP applications, it can be difficult to set up an EEG cap, especially with a higher number of electrodes.

Normally, SSVEP signals are collected from the visual cortex at locations $O1$, $O2$ and $Oz$ (based on the 10–20 international standard), since these locations have the highest response for visual stimulus [13]. Figure 1 shows a functional representation of a basic BCI system. Previous studies have explored EEG data collected from multiple electrodes at various locations in and around the ear region [14–16] to circumvent issues related to the cumbersome scalp electrodes. A recent study also explored the possibilities of acquiring SSVEP data behind the ear areas using various types of visual stimuli [17].

Recently, behind the ear EEG data collection was also explored without an external stimuli based on cognitive tasks demonstrating that ear area could be used for developing BCI system which could be further extended using external stimuli [18].

Considering a wider range of commercially available EEG hardware as discussed above which are either expensive or complex to set up for controlling external applications using SSVEP, in-ear EEG collection using 15 electrodes in each ear was explored in a previous study which required a complex hardware (difficult considering a practical BCI application) [14]. Study based on six ear electrodes in each ear investigated the possibilities of collecting EEG identified the signal-to-noise ratio is closer to that of scalp was also not an easier solution for a portable BCI system [19]. We propose this single electrode system, which can be easily fitted like an earphone and EEG data could be acquired. Acquiring SSVEP data using a single in-ear electrode, makes practical setup easier for daily use.

We developed hardware and electronics and experiment with various visual stimulus frequencies, comparing results against SSVEP data acquired from a traditional scalp EEG in the occipital region.

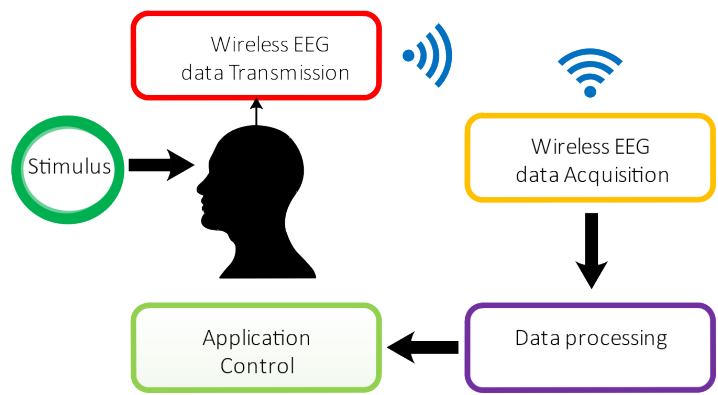

**Figure 1.** BCI data acquisition and control block diagram.

## 2. Methodology

To investigate the feasibility of acquiring reliable SSVEP signal via the ear canal, conductive rubber from Estim with a diameter of 3–4 mm was used as an electrode, shaped as a disposable ear plug as shown in Figure 2. The conductive rubber had very low resistance of approximately 5–10 Ω, for a length of 10–12 mm. The rubber electrode was connected to a standard electrode cable, wired to a standard EEG amplifier and acquisition system.

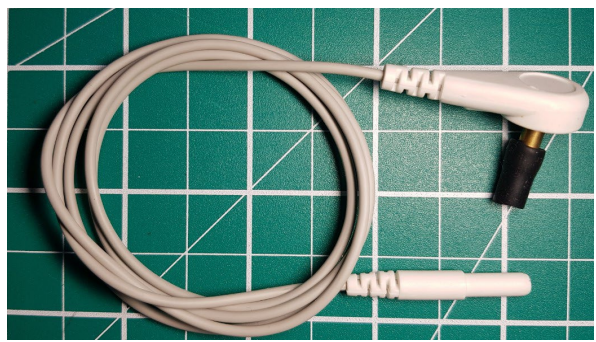

**Figure 2.** Conductive rubber electrode.

### 2.1. Hardware Design

To generate precise flicker frequencies, an ARM 32-bit microcontroller based on a Teensy development platform was used [20]. The frequencies used where 7, 9, 11 and 13 Hz with a duty-cycle of 50%. Relatively low frequencies were chosen for this study since they are known to yield reliable SSVEP amplitudes [21] to aid comparison between in-ear and scalp electrodes. We deliberately did not present the stimulus using an LCD screen or computer display since they are restricted by screen refresh rates. Instead, flickers were generated using a green chip-on-board (COB) circular LED ring with a diameter of 130 mm as shown in Figure 3; this configuration was found to have the highest SSVEP performance in a previous study using scalp electrodes [22]. The hardware schematic for the flicker generator is shown in Figure 4, showing the 72 MHz Cortex-M4 based controller, which can generate very precise flicker signals. The firmware was programmed individually for the required frequencies, and the digital control signal fed to a MOSFET for driving the LED ring at constant brightness. An external 12 V DC source was used through a high current DC-DC converter to avoid any external power line interference. To simultaneously generate multiple frequencies and to drive additional visual stimuli, it is possible to cascade the modules shown in the schematic of Figure 4.

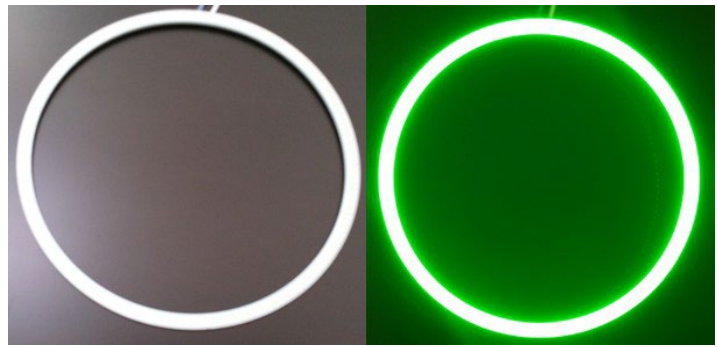

**Figure 3.** The chip-on-board radial LED unlit (**left**) and lit (**right**).

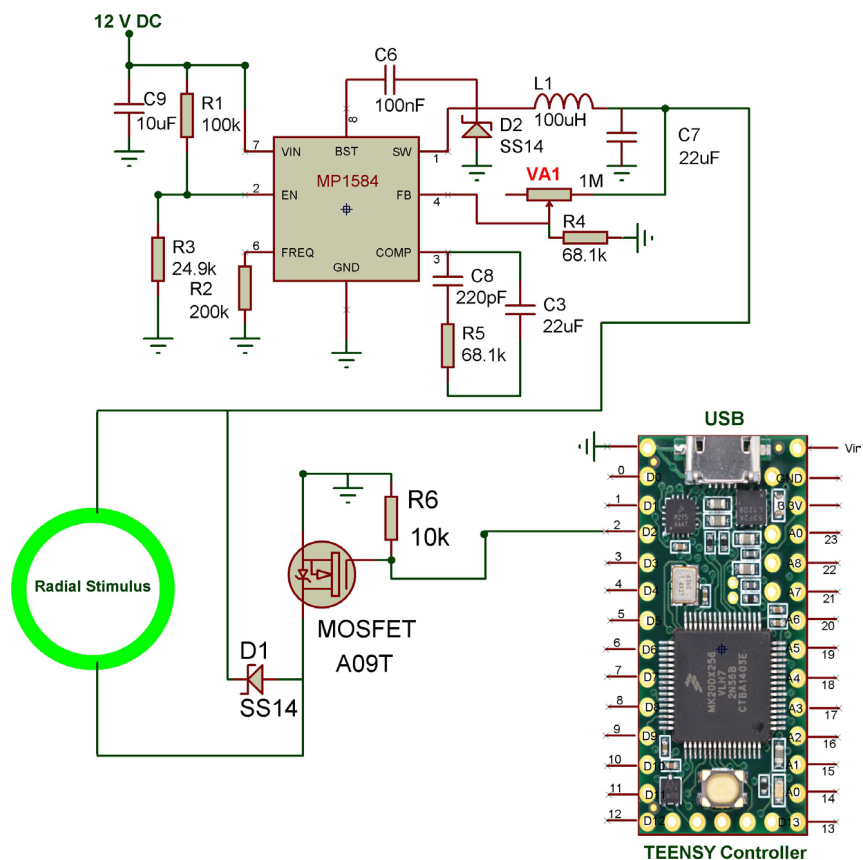

**Figure 4.** Visual stimulus generator circuit schematic.

## 2.2. Data Acquisition

For this study, EEG was recorded using Open BCI wireless EEG acquisition hardware with a sampling rate of 250 Hz. The layout for the electrode is shown in Figure 5. Five participants volunteered for this study, having perfect or corrected vision (three males, two females, with age $34 \pm 12.1$ years). Two standard scalp electrodes, $O1$ and $O2$, were fitted on the EEG cap, and a third conductive rubber electrode was inserted inside the ear canal (which was previously cleaned with alcohol wipes to remove wax), and located at a safe distance from the ear drum. Conductive gel was applied to the scalp electrodes to ensure proper conductivity, but not to the in-ear electrode. Experimental setup involved participants being comfortably seated, the electrodes installed, and then the visual stimulus placed a distance of 60 cm away at eye level. All the experiments carried out with human participants were in accordance with The Code of Ethics of the World Medical Association (Declaration of Helsinki).The study received prior ethical approval from the University of Kent Faculty of Sciences Ethics Committee (Project Identification Reference: 0021516) and all participants gave written consent before participating.

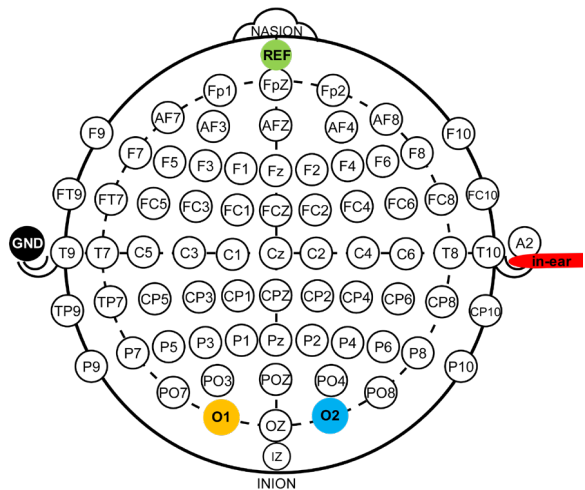

**Figure 5.** Electrode layout for EEG data acquisition.

The visual stimulator was programmed with the desired frequencies to evoke SSVEP for a period of 30 s for each trial. EEG recording sessions started with one frequency trial for 30 s, then participants were given a break of one minute to allow any previous stimulus influences to subside before the next trial commenced. Five 30-second trials were completed for each frequency, and were presented in random order. All of the recorded data was anonymised and stored for further analysis.

## 2.3. Data Analysis

The data collected from both occipital and in-ear electrodes were analysed in MATLAB. For each frequency, the EEG data from channels $O1$, $O2$ and the in-ear electrode were filtered using a fourth order Infinite Impulse Response (IIR) Elliptical filter with bandpass range from 6 to 14 Hz with stop band edges at 5 and 15 Hz. The extracted 30-second data was normalised in the range [0,1] since the in-ear data was of higher amplitude than the scalp data. A section of the EEG signal acquired simultaneously from an occipital and in-ear electrode is shown in Figure 6. Spectrograms of the data were formed (in the wide band frequency range 5 Hz–40 Hz) to explore the visual stimulus response for the different electrode locations. The wide band filter was used to show the similarity of scalp and in-ear electrodes in general,i.e. not for any specific frequency range.

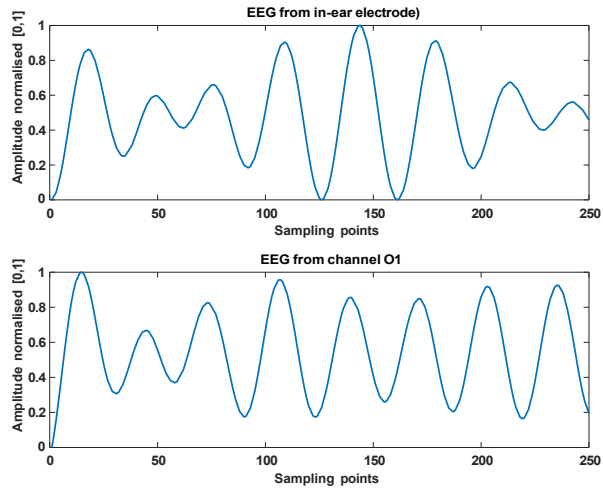

**Figure 6.** A one second section of EEG waveform acquired simultaneously from the in-ear electrode and occipital channel O1 for stimulus frequency of 7 Hz (filtered from 6 to 8 Hz to clearly show the cyclic information).

The two occipital channels were averaged prior to analysis. To visually assess the quality of the data from both locations, Figures 7–11 compare a Section (15 s, as an example) spectrograms from the occipital and in-ear electrode locations. Plots are presented from all five participants for every stimulus frequency, and both recording locations.

The data was also analysed using Fast Fourier Transforms (FFT). For FFT analysis, each 30-s data was segmented into one-second EEG segments and analysed with Fast Fourier Transform (FFT). The maximum amplitudes of the FFT for all 150 segments from five trials for each frequency were computed and stored separately for each participant.

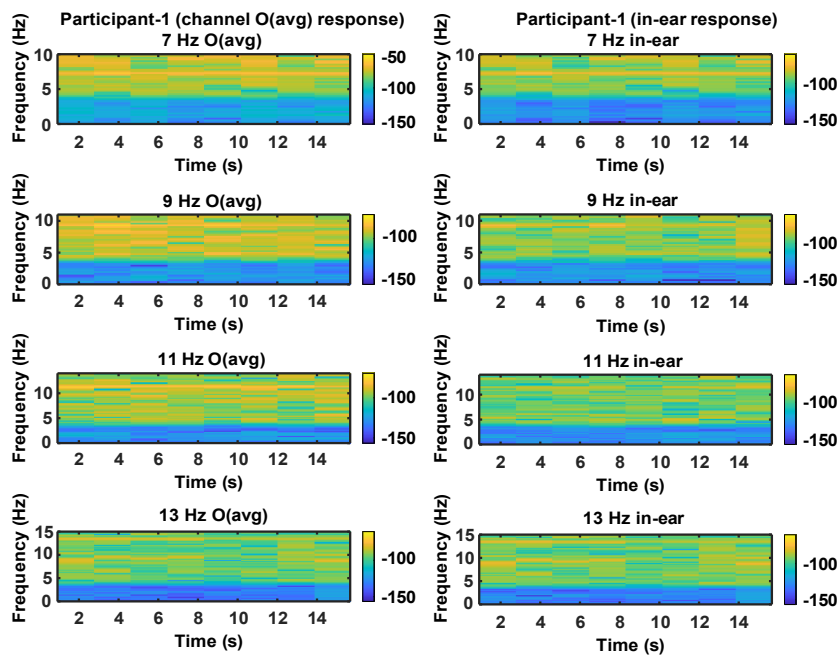

**Figure 7.** Spectrogram of participant 1 from occipital and in-ear electrode.

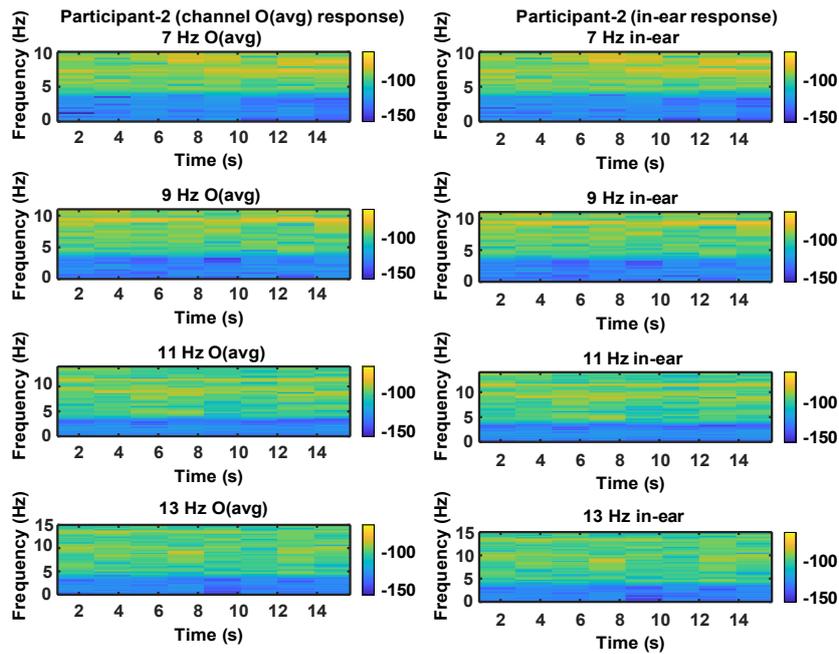

**Figure 8.** Spectrogram of participant 2 from occipital and in-ear electrode.

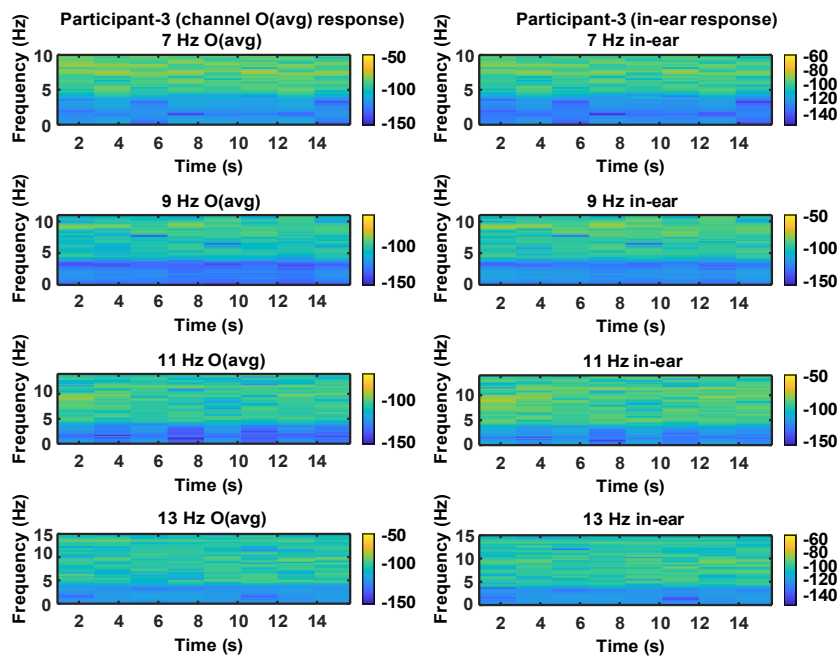

**Figure 9.** Spectrogram of participant 3 from occipital and in-ear electrode.

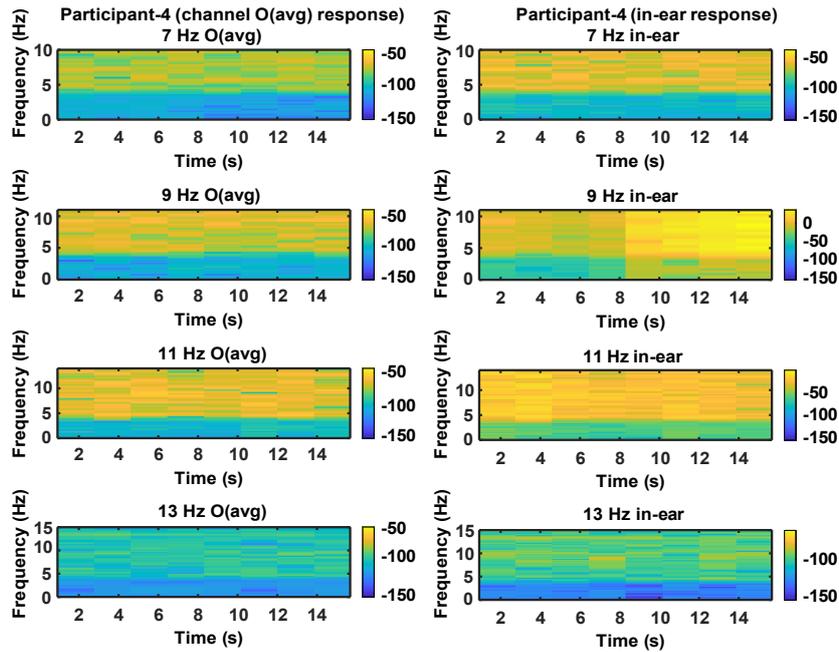

**Figure 10.** Spectrogram of participant 4 from occipital and in-ear electrode.

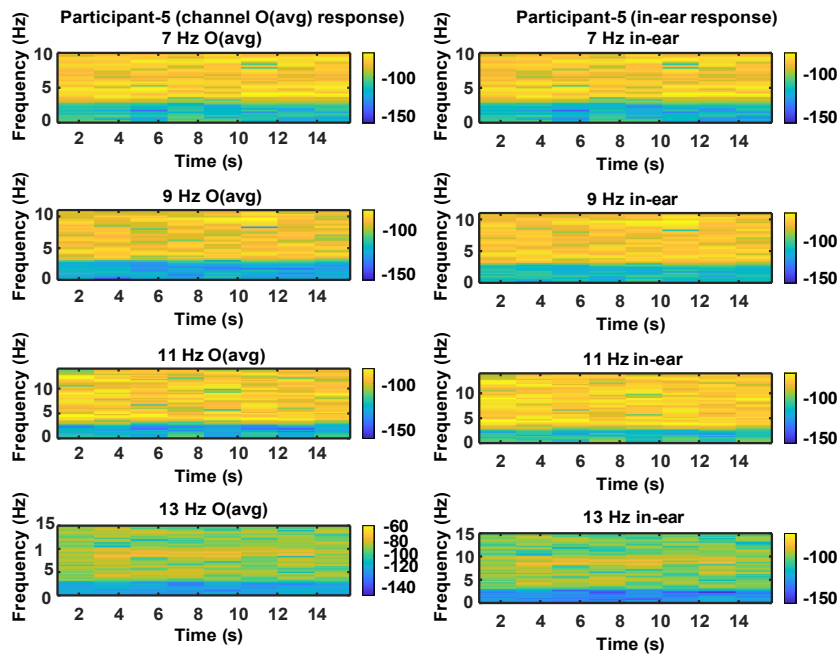

**Figure 11.** Spectrogram of participant 5 from occipital and in-ear electrode.

## 3. Results and Discussion

As mentioned earlier, in-ear and occipital SSVEP responses are compared for signal reliability. Four frequency visual stimuli were used to collect the SSVEP EEG responses from five participants. From the spectrograms (extracted for a sample of 15 s of EEG), it can be observed that, the stimulus frequency extracted from the occipital region is comparable to the signal extracted from the in-ear electrode. From Figure 6 also it can be seen that both occipital and in-ear has similar cyclic pattern in the acquired data. The figure shows a section of the EEG - one second extract which has been filtered from 6 to 8 Hz — to more clearly indicate the cyclic information present in both the in-ear and occipital EEG for stimulus frequency of 7 Hz.

To compute the magnitude of association between occipital and in-ear responses, Pearson's correlation coefficient (*r*) along with *p*-values, were computed for all the stimulus frequencies combining all participant data for each frequency.

Signal to noise ratio (SNR) for each frequency was computed by the magnitude of FFT of the specific frequency over average of the magnitude of FFT of three other frequencies. For example, SNR for 7 Hz was computed using:

$$SNR_7 = 20 * log_{10} \frac{|FFT_7|}{\frac{1}{3}\sum_{i=9,11,13} |FFT_i|} \quad (1)$$

The bandwidths were computed as the width when the power of the peak frequency dropped by 3 dB. Peak frequency, SNR and bandwidth for all the stimuli frequencies from five participants are shown in Table 1. The SNR values were always positive showing that the SSVEP peak frequency does match the flashing frequency. However, the peak frequency is not always exactly at the required frequency, which could be due to noise and issue with the computation of the frequency (spectral leakage). However, in all the cases the peak frequency values were in ±0.5 Hz of the expected SSVEP frequency. The bandwidth computations showed that these values were within less than 1 Hz which again indicated good SSVEP response.

**Table 1.** Peak frequency, SNR and bandwidth of elicited SSVEP responses.

| Participant | Stimulus Flicker Frequency (Hz) | Peak Frequency (Hz) | | SNR (dB) | | Bandwidth (Hz) | |
|---|---|---|---|---|---|---|---|
| | | O (Avg) | In-Ear | O (Avg) | In-Ear | O (Avg) | In-Ear |
| P1 | 7 | 7.10 | 7.10 | 0.02 | 7.51 | 0.46 | 0.10 |
| P2 | 7 | 7.10 | 7.10 | 8.57 | 12.63 | 0.10 | 0.14 |
| P3 | 7 | 7.10 | 7.42 | 3.92 | 27.84 | 0.10 | 0.14 |
| P4 | 7 | 7.10 | 7.10 | 1.19 | 4.20 | 0.21 | 0.28 |
| P5 | 7 | 7.10 | 7.42 | 3.75 | 8.57 | 0.10 | 0.17 |
| Avg ± Std | | 7.1 ± 0 | 7.22 ± 0.17 | 3.49 ± 3.29 | 12.15 ± 9.27 | 0.19 ± 0.15 | 0.16 ± 0.06 |
| P1 | 9 | 9.10 | 9.10 | 4.96 | 3.02 | 0.17 | 0.21 |
| P2 | 9 | 9.14 | 9.12 | 3.80 | 3.86 | 1.14 | 1.32 |
| P3 | 9 | 9.03 | 8.96 | 1.78 | 16.25 | 0.53 | 0.14 |
| P4 | 9 | 9.14 | 9.16 | 11.06 | 11.06 | 0.21 | 11.55 |
| P5 | 9 | 9.35 | 9.46 | 9.39 | 7.11 | 0.14 | 0.35 |
| Avg ± Std | | 9.15 ± 0.11 | 9.16 ± 0.18 | 6.19 ± 3.89 | 8.26 ± 5.47 | 0.44 ± 0.42 | 2.71 ± 4.96 |
| P1 | 11 | 11.03 | 11.10 | 3.69 | 0.14 | 0.14 | 0.21 |
| P2 | 11 | 11.10 | 11.11 | 6.06 | 0.59 | 0.28 | 0.10 |
| P3 | 11 | 11.11 | 11.10 | 5.56 | 0.72 | 0.29 | 0.43 |
| P4 | 11 | 11.10 | 10.56 | 1.89 | 1.87 | 0.35 | 0.21 |
| P5 | 11 | 11.25 | 11.45 | 1.68 | 5.40 | 0.18 | 0.11 |
| Avg ± Std | | 11.11 ± 0.08 | 11.06 ± 0.31 | 1.62 ± 2.02 | 1.74 ± 2.14 | 0.24 ± 0.08 | 0.21 ± 0.13 |
| P1 | 13 | 13.07 | 13.10 | 1.76 | 1.09 | 0.25 | 0.35 |
| P2 | 13 | 13.1 | 13.07 | 1.09 | 0.39 | 0.35 | 0.17 |
| P3 | 13 | 13.11 | 13.10 | 5.87 | 4.77 | 0.28 | 0.35 |
| P4 | 13 | 13.07 | 13.10 | 2.78 | 1.36 | 0.18 | 0.21 |
| P5 | 13 | 13.07 | 13.12 | 2.86 | 3.02 | 0.11 | 0.1 |
| Avg ± Std | | 13.08 ± 0.01 | 13.09 ± 0.01 | 2.87 ± 1.83 | 2.12 ± 1.76 | 0.23 ± 0.09 | 0.23 ± 0.11 |

The spectrograms do not clearly show the SSVEP frequency as the filter is wide band from 5 to 40 Hz but all the figures show similarity between O (ave) and in-ear, which denotes that in-ear EEG rhythms are similar to those obtained in O (ave). Furthermore, from the spectrogram plots, it can be seen the lower frequencies have higher response for both occipital and in-ear, which is the expected situation for SSVEP EEG.

The results show it is possible to replace the scalp electrode with the proposed in-ear EEG electrode for SSVEP based BCI. In-ear electrode could be used as part of a wearable accessory, which would make it more practical for developing support platforms based on BCI.

As mentioned earlier, maximum FFT amplitude for each frequency was computed for all the participants separately. For each frequency, the 150 values from five trials were added for each participant from channels *O1*, *O2* and in-ear. This resulted in 750 values for each frequency—the average of O1 and O1 for all 750 values were computed to be compared with the in-ear FFT amplitude. Figure 12, shows the box plot for the averaged occipital amplitude compared to the in-ear amplitude. For each frequency, the in-ear amplitude is slightly less than the occipital amplitude as the occipital region has the highest response for SSVEP.

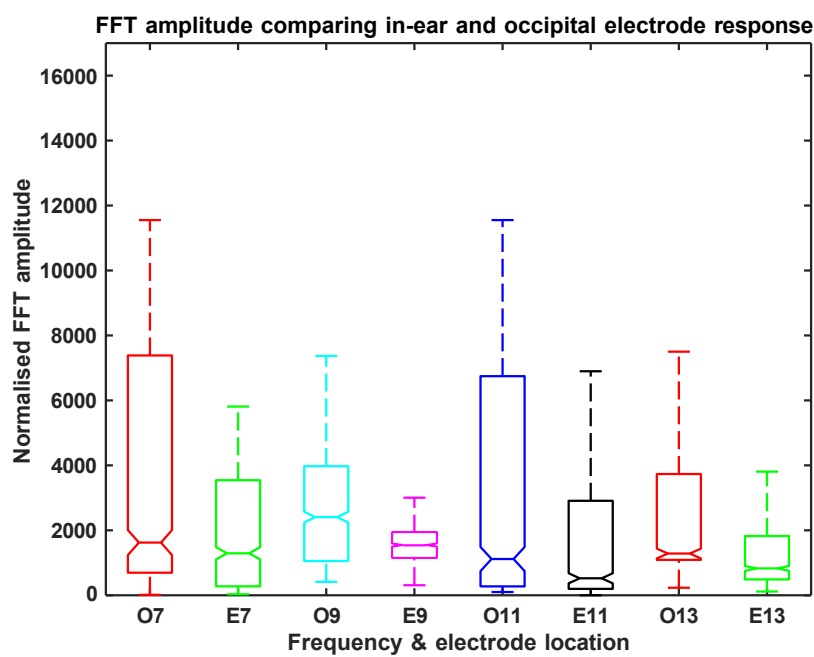

**Figure 12.** Normalised FFT amplitude for occipital and in-ear location.

Figure 13 shows the scatter plot analysis for 9 Hz (as an example) with positive correlation from averaged occipital and in-ear electrodes. Table 2 shows correlation coefficients and *p*-values for all stimulus frequencies used in the experiment. The *p*-values for the computed Pearson's coefficients are below 0.05 indicating significance.

**Table 2.** Pearson's correlation coefficient (N = 750).

| Stimulus Frequency (Hz) | Coefficient (*r*) | *p*-Value (*p*) |
|---|---|---|
| 7 Hz | 0.76 | 0.03 |
| 9 Hz | 0.89 | 0.03 |
| 11 Hz | 0.79 | 0.04 |
| 13 Hz | 0.85 | 0.04 |

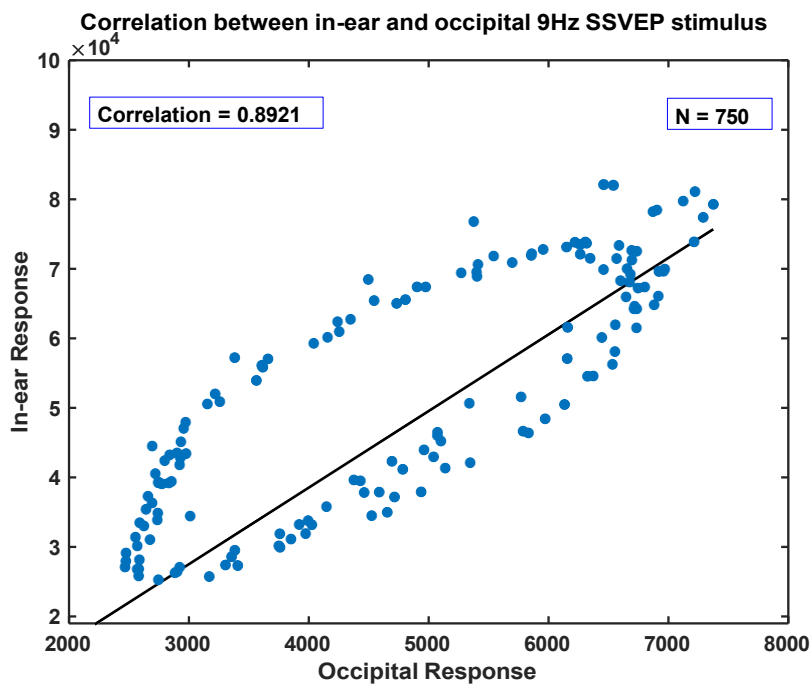

**Figure 13.** Scatter plot showing the correlation between occipital and in-ear SSVEP.

## 4. Conclusions

This paper has explored non-invasive acquisition of EEG data using in-ear electrodes. SSVEP was elicited using 7, 9, 11 and 13 Hz visual stimulus frequencies and acquired simultaneously from occipital and in-ear electrodes.

The SNR values were always positive showing that the SSVEP peak frequency does match the flashing frequency. The computed bandwidth showed that these values were within less than 1 Hz which also indicated good SSVEP response. Spectrogram data from both locations also show similarity between O (ave) and in-ear, which denotes the EEG rhythms are similar to those obtained in O (ave). This is further supported by Pearson's correlation, which also indicates that it is possible to use in-ear EEG data for SSVEP based BCI. This clearly supports the potential of in-ear electrodes in future to serve as a reliable source for SSVEP data acquisition.

Wearable SSVEP based BCI applications would clearly benefit in terms of usability and practicality if they adopted the proposed in-ear EEG electrodes. Future studies will explore performance under noisy environments such as in the presence of frequent eye blinks, during movement and in real-world conditions (with increased electrical interference). We believe in-ear electrodes have potential to be less susceptible to such interference compared to scalp electrodes, which would further enhance their practicality.